\documentclass[conference]{IEEEtran}
\IEEEoverridecommandlockouts
\usepackage[ansinew]{inputenc}
\usepackage[pdftex]{graphicx}
\usepackage[T1]{fontenc}
\usepackage{color,amsmath,enumerate,amssymb,hhline,verbatim}

\begin{document}

\title{A Connection Between Locally Repairable Codes and Exact Regenerating Codes}

\author{
\IEEEauthorblockN{Toni Ernvall\IEEEauthorrefmark{1}, Thomas Westerb\"ack\IEEEauthorrefmark{1}, Ragnar Freij-Hollanti\IEEEauthorrefmark{2}, and Camilla Hollanti\IEEEauthorrefmark{1}}
\IEEEauthorblockA{
\IEEEauthorrefmark{1}Department of Mathematics and Systems Analysis, Aalto University, P.O.Box 11100, FI-00076 Aalto, Finland\\
Emails: \{firstname.lastname\}@aalto.fi}
\IEEEauthorblockA{\IEEEauthorrefmark{2}Department of Communications and Networking, Aalto University, P.O.Box 13000, FI-00076 Aalto, Finland\\
Email: ragnar.freij@aalto.fi}
\thanks{This work was financially supported by the Academy of Finland grants \#276031, \#282938, and \#283262 to C. Hollanti, and by a grant to C. Hollanti from Magnus Ehrnrooth Foundation, Finland. The support from the European Science Foundation under the ESF COST Action IC1104 is also gratefully acknowledged.}
}

\maketitle

\newtheorem{definition}{Definition}[section]
\newtheorem{thm}{Theorem}[section]
\newtheorem{proposition}[thm]{Proposition}
\newtheorem{lemma}[thm]{Lemma}
\newtheorem{corollary}[thm]{Corollary}
\newtheorem{exam}{Example}[section]
\newtheorem{conj}{Conjecture}
\newtheorem{remark}{Remark}[section]

\newcommand{\La}{\mathbf{L}}
\newcommand{\h}{{\mathbf h}}
\newcommand{\Z}{{\mathbf Z}}
\newcommand{\R}{{\mathbf R}}
\newcommand{\C}{{\mathbf C}}
\newcommand{\D}{{\mathcal D}}
\newcommand{\F}{{\mathbf F}}
\newcommand{\HH}{{\mathbf H}}
\newcommand{\OO}{{\mathcal O}}
\newcommand{\G}{{\mathcal G}}
\newcommand{\A}{{\mathcal A}}
\newcommand{\B}{{\mathcal B}}
\newcommand{\I}{{\mathcal I}}
\newcommand{\E}{{\mathcal E}}
\newcommand{\PP}{{\mathcal P}}
\newcommand{\Q}{{\mathbf Q}}
\newcommand{\M}{{\mathcal M}}
\newcommand{\separ}{\,\vert\,}
\newcommand{\abs}[1]{\vert #1 \vert}

\begin{abstract}
Typically, locally repairable codes (LRCs) and regenerating codes have been studied independently of each other, and it has not been clear how the parameters of one relate to those of the other. In this paper, a novel connection between locally repairable codes and exact regenerating codes is established. Via this connection, locally repairable codes are interpreted as exact regenerating codes. Further, some of these codes are shown to perform better than time-sharing codes between minimum bandwidth regenerating and minimum storage regenerating codes.
\end{abstract}

\section{Introduction}

In the field of distributed storage, regenerating codes and locally repairable codes (LRCs) are two important classes of codes. These two code types are designed for different purposes. Namely, a LRC is primarily constructed to have small sets of repair nodes, whereas regenerating codes require every set of a predetermined size to be able to repair a node. Naturally, the question arises as to what kind of a regenerating code does a LRC imply. Our aim is to answer this question by  illustrating a fundamental connection between the two code types.

\subsection{Exact Regenerating Codes}
\emph{An exact regenerating code} \cite{explicitconst,reducingrepair} with parameters $(n,k,d)$ with node size $\alpha$, total repair bandwidth $\gamma$, and file size $B$ is a system with $n$ storage devices (called \emph{nodes}) with the property that any $k$ nodes can recover the stored file and any $d$ nodes (called \emph{helper nodes}) can repair a lost node with an exact copy by transmitting a total amount $\gamma$ of information. If each helper node transmits the same amount $\beta$ of information, we say that a system has \emph{symmetric repair}. In that case we have $\gamma=d\beta$.

The concept of \emph{functionally repairing codes} was introduced in \cite{dimakis}. In that scenario we do not require the replacing node to be an exact copy of the old one. For functionally repairing codes it was proved in the same paper that the possible values of $B$ are characterized by
\begin{equation}\label{eq:dimakiscapacity}
B \leq \sum_{j=0}^{k-1} \min\{ \alpha , (d-j)\beta \}.
\end{equation}
Since exact repair is a stronger requirement than functional repair, the same bound holds for exact regenerating codes too. The bound (\ref{eq:dimakiscapacity}) defines a tradeoff between the node size and the total repair bandwidth. The point on the tradeoff curve in which $\gamma$ is minimized is called the \emph{minimum bandwidth regenerating} (MBR) point
\[
\begin{cases}
    \alpha=\frac{2dB}{k(2d-k+1)}  \\
    \gamma=\frac{2dB}{k(2d-k+1)}. \\
\end{cases}
\]
The point in which $\alpha$ is the smallest possible is called the \emph{minimum storage regenerating} (MSR) point
\[
\begin{cases}
    \alpha=\frac{B}{k}  \\
    \gamma=\frac{dB}{k(d-k+1)}. \\
\end{cases}
\]
These extremal points are known to be achievable at least asymptotically also under the assumption of exact repair \cite{kumar,yhteensaMDS}.

\subsection{Locally Repairable Codes}
Locally repairable codes were first studied in \cite{OnTheLocality}, \cite{SelfRepairHomomorphic}, \cite{LRCpapailiopoulos}. A locally repairable code with parameters $(N,K,D,R,\Delta)$ is a code over a finite field of size $q$ with length $N$,  cardinality  $q^K$, minimum distance $D$, and  $(R,\Delta)$ all-symbol locality property meaning that each symbol has $(R,\Delta)$ locality. A symbol $j \in \{1,\dots,n \}$ has $(R,\Delta)$ locality if there exists a subset $S$ of nodes such that $|S| \leq R+\Delta-1$, $j \in S$ and any node in $S$ can be repaired by any $|S|-\Delta+1$ nodes from $S$. The definition of $(R,\Delta)$ all-symbol locality was introduced in \cite{LocalRegeneration}.

\begin{remark}
Usually the letters $(n,k,d,r,\delta)$ are used to denote the parameters of a LRC. Here, we use the capital letters $(N,K,D,R,\Delta)$ to distinguish the parameters of a LRC from the parameters $(n,k,d)$ of a regenerating code.
\end{remark}

In \cite{LocalRegeneration}, the following upper bound was derived on the minimum distance of linear LRCs,
\begin{equation}\label{eq:minDistanceUpperBound}
D \leq N-K-\left( \left\lceil\frac{K}{R}\right\rceil -1\right)(\Delta-1)+1.
\end{equation}
In \cite{Rawat}, the same bound was later established for nonlinear LRCs.

\subsection{Contributions and Related Work}
In Section \ref{Subsec:connection} we introduce a fundamental connection between LRCs and exact regenerating codes. To the best of our knowledge, this connection has not been pointed out before. The connection gives a method to use LRCs as exact regenerating codes. In Section \ref{Sec:Construction}, we give a method to construct LRCs that have good properties when interpreted as exact regenerating codes. In addition, we will analyze the performance of such exact regenerating codes in Section \ref{Sec:Performance}. 

Optimal exact regenerating codes at the MBR point for all values of $(n,k,d)$ and optimal exact regenerating codes at the MSR point when $d \geq 2k-2$ are constructed in \cite{kumar}. The asymptotic achievability of the MSR point for all values of $(n,k,d)$
is shown in \cite{yhteensaMDS}.

Constructions of exact regenerating codes between the MBR and MSR points exceeding the time-sharing line are studied in \cite{exactrepairtian,sasidharan,exactjournal,goparaju}. Results in \cite{exactrepairtian} and \cite{sasidharan} are combined in \cite{TianSasidharan}. However, in contrast to these constructions we are not trying to find codes that perform as well as possible. Instead, our main purpose is to show that LRCs can be used as exact regenerating codes and, in many cases, they have quite good performance. However, when compared to the other constructions, LRCs when used as exact regenerating codes do not usually perform that well. The established connection also enables further analysis on LRCs in terms of storage space, repair bandwidth, \emph{etc}.

From the opposite point of view, nontrivial upper bounds for exact regenerating codes are studied for example in \cite{outerboundSasidharan} and \cite{duursmaOuterBound}.
 In the case $(n,k,d)=(4,3,3)$, the capacity curve for all possible values of $(\alpha,\gamma)$ was solved in \cite{region433}. In the nonasymptotic case, it was shown in \cite{nonachievability} that in most points between MBR and MSR, there do not exist exact regenerating codes that would achieve the capacity of functionally repairing regenerating codes.

Locally repairable codes for several parameter sets $(N,K,D,R,\Delta)$ are constructed in \cite{OnTheLocality}, \cite{LocalRegeneration}, \cite{LRCmatroid}, \cite{SongOptimal}, \cite{Rawat}, \cite{matroidLRC}. In \cite{OnTheLocality} it was proved that bound \eqref{eq:minDistanceUpperBound} is not achievable for all parameters $(N,K,R)$. Codes with local regeneration are studied in \cite{LocalRegeneration}. In a completely different direction, connections between regenerating codes and LRCs have been made in \cite{Ahmad}, where local repairability is exploited to select helpers in a regenerating code.

\section{Connection Between Locally Repairable Codes and Exact Regenerating Codes}\label{Subsec:connection}
The next theorem describes an interesting connection between LRCs and exact regenerating codes.
\begin{thm}\label{thm:connection}
Suppose we have a LRC of length $N$, dimension $K$, minimum distance $D$ and $(R,\Delta)$ all-symbol locality. The same code can be regarded as an exact regenerating code with length $n=N$, file recover degree $k=N-D+1$, node repair degree $d=N-\Delta+1$, node size $\alpha=1$, total repair bandwidth $\gamma=R$, and file size $B=K$. In addition, there exists an exact regenerating code with symmetric repair for the same parameters $n$, $k$ and $d$ with node size $\alpha'=n!$, total repair bandwidth $\gamma'=n!R$ and file size $B'=n!K$.
\end{thm}
\begin{IEEEproof}
Consider a LRC with parameters $(N,K,D,R,\Delta)$. Clearly, its length is $n=N$ and node size is $\alpha=1$.  By the definition of minimum distance, by contacting any $N-D+1$ nodes we may recover the original file.

Suppose we have lost a node that belongs to locality group $S_j$. Contacting any $N-\Delta+1$ of the existing nodes is enough to guarantee that we have at least $|S_j|-\Delta+1$ nodes from the set $S_j$. Hence, we are able to repair the lost node.
As the lost node could be repaired by $R$ nodes from $S_j$, the total repair bandwidth is $\gamma=R$.

In general, the above code does not have symmetric repair. However, taking $n!$ copies of the code and rearranging the order of the nodes according to each permutation on the set $\{1,\dots,n\}$, and then finally storing onto $n$ nodes we get a code with symmetric repair. This method is explained in more detail in \cite{Heterogeneous}.
\end{IEEEproof}

\begin{remark}
In Theorem \ref{thm:connection}, after normalization, the exact regenerating code with symmetric repair has node size $\frac{\alpha'}{n!}=1$, total repair bandwidth $\frac{\gamma'}{n!}=R$, and file size $\frac{B'}{n!}=K$.
\end{remark}

\begin{remark}
The constructions of \cite{exactjournal} have an important similarity with Theorem \ref{thm:connection}: in both cases the number of helper nodes used in the resulting exact regenerating code is bigger than in the corresponding LRC. This property is used in both cases to obtain that any set of certain size is able to repair a lost node.
\end{remark}

Notice that when using a LRC, we transmit all the stored information when repairing a lost node. Thus, while the definition of regenerating codes allows us to send only parts of the stored information, in our derived regenerating codes we do not modify the information before transmitting it.

Define
\[
C_{n,k,d}^{LRC}(1,\gamma)
\]
to be the largest file size (that is, the largest possible $K$) that can be stored using a LRC with $N=n$, $R \leq \gamma$, $D \geq n-k+1$ and $\Delta \geq n-d+1$. Notice that by taking $\alpha$ copies of the system we can store a file that is of size $\alpha$ times the original file size. Hence, we define generally
\[
C_{n,k,d}^{LRC}(\alpha,\gamma)=\alpha C_{n,k,d}^{LRC}(1,\frac{\gamma}{\alpha}).
\]
Also, for functionally-repairing regenerating codes, denote the capacity by
\[
C_{k,d}(\alpha,\gamma) = \sum_{j=0}^{k-1} \min \left\{ \alpha , \frac{(d-j)\gamma}{d} \right\}.
\]
 Clearly, $C_{n,k,d}^{LRC}(\alpha,\gamma) \leq C_{k,d}(\alpha,\gamma)$.

Remarkably, these na\"ively constructed exact regenerating codes have good properties, when the parameters are chosen wisely. In Section \ref{Sec:Performance}, we will give an asymptotic analysis of the behavior of $C_{n,k,d}^{LRC}(\alpha,\gamma)$ when $n$, $k$ and $d$ are large and close to each other.

\begin{remark}
Several variations of the definition of  a LRC are studied in the literature. For example, in \cite{multiplerepair} and \cite{TamoOptimalLRC}, 
codes with multiple repair alternatives are studied. Similarly as before, such a code could be used as an exact regenerating code.
\end{remark}

\section{Construction}\label{Sec:Construction}
In most of the existing literature, LRCs are constructed with the parameters $N$, $K$, $R$ and $\Delta$ fixed, and the minimum distance $D$ as large as possible. In this paper, however, we want to find an exact regenerating code with as good performance as possible when $n$, $k$, $d$, $\alpha=1$ and $\gamma$ are fixed. This means that $N=n$, $D=n-k+1$, $R=\gamma$ and $\Delta=n-d+1$ are fixed and we need to find a LRC with best possible code dimension $K$. Generally, the largest possible value of $K$ is not known. In this section we give a method for finding LRCs for this purpose. 

The following proposition for linear LRCs follows immediately from Theorem V.5 (i) and (ii) in \cite{matroidLRC}. 

\begin{proposition} \label{prop:construction}
Let $(N,D,R,\Delta)$ be integers such that $1 \leq R$, $2 \leq \Delta \leq D \leq N - R + 1$ and $s = \lceil \frac{N}{R + \Delta - 1}\rceil (R + \Delta - 1) - N$. Then the following hold.
$$
\begin{array}{rl}
(i) & \hbox{If $s <R$, then for sufficiently large $q$ there is a linear} \\
    & \hbox{$(N,K,D,R,\Delta)$-LRC over $\mathbb{F}_q$ with}\\
    & \hbox{$K = R(i+1) - s -t = N + 1 - D - (\Delta - 1)i$,}  \\
    & \hbox{where $i \geq 0$ and $0 \leq t < R$.}\\
(ii) & \hbox{If $s \geq R$, then for sufficiently large $q$ there is a linear} \\
    & \hbox{$(N,K,D,R,\Delta)$-LRC over $\mathbb{F}_q$ with}\\
    & \hbox{$K = R(i+1) - t = $}\\
    & \hbox{$N + 1 - D - (\Delta - 1)i - (R+\Delta-1-s)$, }  \\
    & \hbox{where $i \geq 1$ and $0 \leq t < R$.}
\end{array}
$$
\end{proposition}

We will here describe how a $K \times N$ generator matrix over a sufficiently large finite field $\mathbb{F}_q$ can be constructed for a linear LRC obtained in Proposition \ref{prop:construction}. First, index the columns of $G$ from 1 to $N$, and let $G_X$ denote the columns indexed by $X$ in $G$, where $X \subseteq \{1,\ldots,N\}$. Further, let $F_1, \ldots,F_m$ denote a partition of $\{1,\ldots,N\}$.

In the case (i), let $m = \lceil \frac{N}{R + \Delta - 1}\rceil$, 
$$
F_j = \{(R + \Delta - 1)(j - 1) + 1, \ldots, (R + \Delta - 1)j\} 
$$
if $1 \leq j \leq m-1$, and 
$$
F_m = \{(R + \Delta - 1)(m - 1) + 1, \ldots, (R + \Delta - 1)m - s\}. 
$$
Now we can construct a a generator matrix $G$ over a sufficiently large $\mathbb{F}_q$ such that
$$
\begin{array}{rl}
(i) & \hbox{$G_{F_j}$ generates a linear maximum distance separable}\\ 
& \hbox{(MDS) code of rank $R$ for $1 \leq j \leq m-1$,}\\
(ii) & \hbox{$G_{F_m}$ generates a linear MDS-code of rank $R-s$},\\
(iii) & \hbox{for any subset $X \subseteq \{1,\ldots,N\}$, the column vectors}\\
      & \hbox{of $G$ indexed by $X$ are linearly independent if and }\\
      & \hbox{only if $|X| \leq K$, $|X \cap F_j| \leq R$ for $1 \leq j \leq m-1$ } \\
      & \hbox{and } |X \cap F_m| \leq R-s. 
\end{array}
$$   
The matrix $G$ generates a linear $(N,K,D,R,\Delta)$-LRC given by the case (i).

In the case (ii), let $m = \lceil \frac{N}{R + \Delta - 1}\rceil - 1$, 
$$
F_j = \{(R + \Delta - 1)(j - 1) + 1, \ldots, (R + \Delta - 1)j\} 
$$
if $1 \leq j \leq m-1$, and 
$$
F_m = \{(R + \Delta - 1)(m - 1) + 1, \ldots, (R + \Delta - 1)(m+1) - s\}. 
$$
Now we can construct a generator matrix $G$ over a sufficiently large $\mathbb{F}_q$ such that
$$
\begin{array}{rl}
(i) & \hbox{$G_{F_j}$ generates a linear MDS-code of rank $R$ for}\\
    & 1 \leq j \leq m,\\
(ii) &  \hbox{for any subset $X \subseteq \{1,\ldots,N\}$, the column vectors}\\
      & \hbox{of $G$ indexed by $X$ are linearly independent if and }\\
      & \hbox{only if $|X| \leq K$ and $|X \cap F_j| \leq R$ for $1 \leq j \leq m$.} 
\end{array}
$$   
The matrix $G$ generates a linear $(N,K,D,R,\Delta)$-LRC given by the case (ii).

We remark that every storage node in a linear LRC is associated with a column in an associated generator matrix. In the above constructions, any node $x \in F_j$ can be repaired by any subset of $R$ nodes in $F_j \setminus \{x\}$, when $1 \leq j \leq m-1$. For $x \in F_m$, in the case (i) any subset of $R-s$ nodes in $F_m \setminus \{x\}$ can repair $x$, and in the case (ii) any subset of $|R|$ nodes in $F_m \setminus \{x\}$ can repair $x$. 

\begin{exam}
The following matrix $G$ over $\mathbb{F}_3$ generates a linear LRC with parameters $(N,K,D,R,\Delta) = (8,5,2,3,2)$ obtained in Proposition \ref{prop:construction} (i). The index partition described in the construction above for this $G$ equals $F_1 = \{1,2,3\}$, $F_2 = \{4,5,6\}$, and $F_3 = \{7,8\}$.  
$$
\begin{small}
G=
\begin{tabular}{ |c|c|c|c|c|c|c|c|}
\multicolumn{1}{c}{{1}}&\multicolumn{1}{c}{{2}}&\multicolumn{1}{c}{{3}}&\multicolumn{1}{c}{{4}}&\multicolumn{1}{c}{{5}}&\multicolumn{1}{c}{{6}}&\multicolumn{1}{c}{{7}}&\multicolumn{1}{c}{{8}}\\
\hline
1&0&1&0&0&0&1&2\\
\hline
0&1&1&0&0&0&1&2\\
\hline
0&0&0&1&0&1&1&2\\
\hline
0&0&0&0&1&1&1&2\\
\hline
0&0&0&0&0&0&1&2\\
\hline
\end{tabular}
\end{small}
$$
\end{exam}

We remark that the linear LRCs obtained in Proposition \ref{prop:construction} (i) are optimal if $\lceil \frac{K}{R} \rceil R - K \geq s$. By optimality, we mean that $K$ is as large as it can be depending on the parameters $(N,D,R,\Delta)$. For the other cases in Proposition \ref{prop:construction}, the obtained dimensions $K$ are lower bounds in the sense that larger dimensions can be possibly found for the associated parameters $(N,D,R,\Delta)$. To get a better $K$ in these cases, one can try to use Theorem 13 or Theorem 14 in \cite{matroidPollanen}, or Theorem V.5 (iii)--(v) in \cite{matroidLRC}.

\section{Performance Analysis}\label{Sec:Performance}

In this section we will analyze how LRCs perform as exact regenerating codes when compared to trivial upper and lower bounds of exact regenerating codes, \emph{i.e.}, when compared to the capacity of functionally-repairing regenerating codes and the time-sharing line between the MBR and MSR points. The performance of the code built by time-sharing MBR and MSR codes is
\[
P_{k,d}^{\text{time-share}}(\alpha,\gamma)=s\frac{k(2d-k+1)\alpha}{2d}+(1-s)k\alpha,
\]
where
\[
\gamma=s\gamma_{MBR}+(1-s)\gamma_{MSR}=s\alpha+(1-s)\frac{d\alpha}{d-k+1}.
\]

\begin{thm}\label{thm:LRCvsTimeShare}
Suppose a LRC with parameters $(N,K,D,R,\Delta)$ is used as an exact regenerating code. It performs better than the code we get by time-sharing MBR and MSR codes if and only if
\begin{equation}\label{eq:LRCvsTimeShare}
K > \frac{N-D+1}{2}\left( 1+\frac{R(D-\Delta+1)}{N-\Delta+1} \right).
\end{equation}
\end{thm}
\begin{IEEEproof}
Recall that $n=N$, $k=N-D+1$, $d=N-\Delta+1$, $\alpha=1$, $\gamma=R$, and the file size is $K$.

The performance of the code built by time-sharing MBR and MSR codes in $\gamma=R$ is
\[
P_{k,d}^{\text{time-share}}(1,\gamma) = s\frac{k(2d-k+1)}{2d}+(1-s)k =  k - s\frac{k(k-1)}{2d} 
\]
with
\[
\gamma=s+(1-s)\frac{d}{d-k+1},
\]
\emph{i.e.},
\[
s=\frac{d}{k-1}-\frac{(d-k+1)\gamma}{k-1}.
\]
Together these equations give
\[
\begin{split}
P_{k,d}^{\text{time-share}}(1,\gamma) & = \frac{k}{2}\left(1 + \frac{(d-k+1)\gamma}{d}\right) \\
& = \frac{N-D+1}{2}\left(1 + \frac{(D-\Delta+1)R}{N-\Delta+1}\right)
\end{split}
\]
proving the claim.
\end{IEEEproof}

\begin{exam}
Suppose $(N,K,D,R,\Delta)=(24,16,4,3,2)$. Now, $R+1$ divides $N$ so there are plenty of constructions showing that such code exists, see for example \cite{OnTheLocality}. Now, the right side of inequality (\ref{eq:LRCvsTimeShare}) is
\[
\frac{24-4+1}{2}\left( 1+\frac{3(4-2+1)}{24-2+1} \right) < 16
\]
and hence such a LRC performs slightly better than the code that is constructed by time-sharing.
\end{exam}

Next we will show that in the case that the parameters $n$, $k$ and $d$ are close to each other, the code rate asymptotically coincides for LRCs and functionally-repairing regenerating codes. The codes constructed in \cite{TianSasidharan}, \cite{goparaju} and \cite{exactjournal} also have  a similar property. However, in the same scenario when compared to codes we get by time-sharing MBR and MSR codes, LRCs as exact regenerating codes perform strictly better.

Suppose $n_m=n+m$, $k_m=k+m$, and $d_m=d+m$, where $n$, $k$ and $d$ are fixed. Let us study the situation where $m$ approaches infinity. This illustrates the case where the parameters $n$, $k$ and $d$ are close to each other.

\begin{thm}\label{thm:limits}
Let $s \in [0,1]$ be fixed and let
\[
\gamma_m=s\alpha+(1-s)\frac{d_m\alpha}{d_m-k_m+1}.
\]
We have
\begin{equation}\label{eq:ExactLrcLimit}
\frac{C_{n_m,k_m,d_m}^{LRC}(\alpha,\gamma_m)}{n_m\alpha} \rightarrow
\begin{cases}
    \frac{1}{n-d+1}  & \quad \text{if } s=1 \\
    1            & \quad \text{if } 0 \leq s < 1 \\
\end{cases}
\end{equation}
as $m \rightarrow \infty$.
\end{thm}
\begin{IEEEproof}
Suppose first that $s=1$, so that $\gamma_m=\alpha=1$ for all $m$. Assume we have a sequence of LRCs with parameters $(N_m=n_m,K_m,D_m \geq n_m-k_m+1,\Delta_m \geq n_m-d_m+1,R_m \leq \gamma_m)$ and consider them as exact regenerating codes. Since $\gamma_m=1$, by equation (\ref{eq:minDistanceUpperBound}) we have
\[
\begin{split}
n-k+1 & \leq n_m-K_m-(K_m-1)(n-d)+1 \\
&=n_m-K_m(n-d+1)+n-d+1.
\end{split}
\]
This gives
\begin{equation}\label{eq:LRCrateMBRupperbound}
\frac{K_m}{n_m} \leq \frac{n_m-d+k}{n_m(n-d+1)} \rightarrow \frac{1}{n-d+1}
\end{equation}
as $m \rightarrow \infty$.




Suppose now that $s=0$ and $k=d$. An MDS code of length $n_m$, dimension $k_m$ and minimum distance $n-k+1$ is known to exist if the field size is large enough and it corresponds to a LRC with $(N_m=n_m,K_m=k_m,D_m=n-k+1,R_m=k_m,\Delta_m=n-k+1)$ and it works as an exact regenerating code with parameters $(n_m,k_m,d_m)$ and $\gamma_m$. Also, its rate approaches one as $m \rightarrow \infty$.

Suppose now that $s \neq 0$ or $k \neq d$. In Theorem V.5 in \cite{matroidLRC} it was shown that when given parameters $N$, $K$, $R$ and $\Delta$ with 
\begin{equation}\label{minDistanceRequirements}
0<R<K \leq N-\left\lceil K/R\right\rceil (\Delta-1),
\end{equation}
we have a LRC for these parameters with minimum distance $D$ satisfying
\[
D \geq N-K-\left\lceil K/R\right\rceil(\Delta-1)+1.
\]
Suppose $N_m=n_m$, $\Delta_m=n_m-d_m+1$, $R_m=\lfloor\gamma_m\rfloor$ and
$
K_m=\left\lfloor S_m\lfloor\gamma_m\rfloor \right\rfloor,
$
where
$
S_m=\frac{m+k+d-n}{n-d+\lfloor\gamma_m\rfloor}.
$
It is easy to check that if $m$ is large enough, then the requirements of equation \eqref{minDistanceRequirements} are satisfied, and hence there exists a LRC for these parameters with minimum distance $D$ at least
\begin{equation}
\begin{split}
& N_m-K_m-\left\lceil\frac{K_m}{R_m}\right\rceil(\Delta_m-1)+1 \\
=\,&  n+m-K_m-\left\lceil\frac{K_m}{\lfloor\gamma_m\rfloor}\right\rceil(n-d)+1 \\
\geq \,& n+m- S_m\lfloor\gamma_m\rfloor -(S_m+1)(n-d)+1 \\
=\, & d+m- \frac{(m+k+d-n)(\lfloor\gamma_m\rfloor+n-d)}{n-d+\lfloor\gamma_m\rfloor}+1 \\
= \,& n-k+1.
\end{split}
\end{equation}
Thus, there exists a LRC with parameters $(N_m,K_m,D_m,R_m,\Delta_m)$ such that it can be used as an exact regenerating code with parameters $(n_m,k_m,d_m)$, $\alpha=1$, $\gamma_m$ and file size $K_m$.

Notice first that since $\alpha=1$, we have
\[
\frac{\lfloor\gamma_m\rfloor}{m} \rightarrow \frac{1-s}{d-k+1}
\]
as $m \rightarrow \infty$.

Now, if $s=1$, then we have $\lfloor\gamma_m\rfloor=1$ and hence the rate of this code is
\begin{equation}
\frac{K_m}{n_m}
= \frac{\left\lfloor S_m \lfloor\gamma_m\rfloor \right\rfloor}{n+m}
= \frac{\left\lfloor \frac{k+m+d-n}{n-d+1} \right\rfloor}{n+m}
\rightarrow \frac{1}{n-d+1}
\end{equation}
as $m \rightarrow \infty$. Together with equation (\ref{eq:LRCrateMBRupperbound}) this gives the first part of equation (\ref{eq:ExactLrcLimit}).

Assume now that $s \neq 1$. Clearly the code rate is upper bounded by $1$. Also, we have
\begin{equation*}
\begin{split}
\frac{K_m}{n_m} = \,& \frac{\left\lfloor \frac{(k+m+d-n)\lfloor\gamma_m\rfloor}{n-d+\lfloor\gamma_m\rfloor} \right\rfloor}{n+m} 
\geq \frac{\frac{(k+m+d-n)\lfloor\gamma_m\rfloor}{n-d+\lfloor\gamma_m\rfloor} -1}{n+m} \\
= \,& \frac{(k+m+d-n)\lfloor\gamma_m\rfloor-(n-d+\lfloor\gamma_m\rfloor)}{(n+m)(n-d+\lfloor\gamma_m\rfloor)} 
\rightarrow  1
\end{split}
\end{equation*}
as $m \rightarrow \infty$, giving the second part of equation (\ref{eq:ExactLrcLimit}).

\end{IEEEproof}

For  comparison, it is straightforward to check that
\begin{equation}\label{eq:TimeshareLimit}
\frac{P_{k_m,d_m}^{\text{time-share}}(\alpha,\gamma_m)}{n_m\alpha} \rightarrow 1-\frac{s}{2},
\end{equation}
\begin{equation}\label{eq:FuncCapLimit}
\begin{split}
\frac{C_{k_m,d_m}(\alpha,\gamma_m)}{n_m\alpha} \rightarrow
\begin{cases}
    \frac{1}{2}  & \quad \text{if } s=1 \\
    1            & \quad \text{if } 0 \leq s < 1 \\
\end{cases}
\end{split}
\end{equation}
as $m \rightarrow \infty$.

Hence the rates of functionally-repairing regenerating  codes and codes built from LRCs are asymptotically the same in each point except in the MBR point. If $n=d+1$, then they are the same also in the MBR point. We conclude by the following immediate consequence of Thm. \ref{thm:limits}.
\begin{corollary}\label{cor:capacities}
Suppose that $s \in [0,1]$ is fixed and let
$
\gamma_m=s\alpha+(1-s)\frac{d_m\alpha}{d_m-k_m+1}.
$
We have
\[
\frac{C_{k_m,d_m}(\alpha,\gamma_m)}{C_{n_m,k_m,d_m}^{LRC}(\alpha,\gamma_m)} \rightarrow
\begin{cases}
    \frac{n-d+1}{2}  & \quad \text{if } s=1 \\
    1            & \quad \text{if } 0 \leq s < 1 \\
\end{cases}
\]
and
\[
\frac{P_{k_m,d_m}^{\text{time-share}}(\alpha,\gamma_m)}{C_{n_m,k_m,d_m}^{LRC}(\alpha,\gamma_m)} \rightarrow
\begin{cases}
    \frac{n-d+1}{2}  & \quad \text{if } s=1 \\
    1-\frac{s}{2}            & \quad \text{if } 0 \leq s < 1 \\
\end{cases}
\]
as $m \rightarrow \infty$.
\end{corollary}

\section{Conclusion}
In this paper, we have shown a connection between locally repairable codes and exact regenerating codes. Namely, a LRC can be used as an exact regenerating code and, after a small modification, as an exact regenerating code with symmetric repair. 
We have utilized this connection to construct exact regenerating codes with better performance than time-sharing MBR and MSR optimal codes. 


\end{document}